\documentclass[a4paper,12pt,reqno,superscriptaddress,nofootinbib]{revtex4}
\usepackage[centertags]{amsmath}
\usepackage{amsfonts}
\usepackage{amssymb}
\usepackage{amsthm}
\usepackage{newlfont}
\usepackage{stmaryrd}
\usepackage{mathrsfs}
\usepackage{euscript}
\usepackage{siunitx}
\usepackage{physics}
\usepackage{algorithm2e}
\usepackage{graphicx,subcaption}
\usepackage{graphicx,caption}
\usepackage{enumitem}
\usepackage{natbib}
\usepackage{graphicx}
\usepackage{color}
\usepackage{floatrow}
\usepackage{caption}
\usepackage{hyperref}
\usepackage{hhline}
\usepackage{hyperref}
\usepackage{cleveref}
\usepackage{import}

\theoremstyle{plain}

\theoremstyle{definition}

\theoremstyle{remark}

\DeclareCaptionJustification{justified}{\leftskip=0pt \rightskip=0pt \parfillskip=0pt plus 1fil}

\numberwithin{equation}{section}

\newcommand{\bbR}{{\mathbb R}}

\newcommand{\opunit}{\text{1}\kern-0.22em\text{l}}

\DeclareMathAlphabet{\mathpzc}{OT1}{pzc}{m}{it}

\newcommand{\fig}{Fig.\;}

\newcommand{\eg}{{\it e.g.\;}}

\newcommand{\id}{\textrm{d}}

\usepackage{floatrow}
\usepackage{caption}

\begin{document}
\title{Local detailed balance for active particle models}
\author{Faezeh Khodabandehlou and Christian Maes\\{\it Department of Physics and Astronomy, KU Leuven, Belgium}}
%\author{Christian Maes} \affiliation{Department of Physics and Astronomy, KU Leuven, Belgium}	\email{faezeh.khodabandehlou@kuleuven.be}

\begin{abstract}
Starting from a Huxley-type model for an agitated vibrational mode, we propose an embedding of standard active particle models in terms of two-temperature processes. One temperature refers to an ambient thermal bath, and the other temperature effectively describes ``hot spots,'' {\it i.e.}, systems with few degrees of freedom showing important population homogenization or even inversion of energy levels as a result of activation. That setup admits to quantitatively specifying the resulting nonequilibrium driving, rendering local detailed balance to active particle models, and making easy contact with thermodynamic features.  % such as discussed in the 1st, 2nd, and 3rd Laws.  
In addition, we observe that the shape transition in the steady low-temperature behavior of run-and-tumble particles (with the interesting emergence of edge states at high persistence) is stable and occurs for all temperature differences, including close to equilibrium.    
\end{abstract}
\maketitle
%\tableofcontents

\section{Introduction}

To study Life with  methods and concepts from statistical physics requires conceiving it as a nonequilibrium process.  Among other things, that demands a quantitative description of the departure from equilibrium.    Life processes typically possess a continuously varying amplitude where the biochemical fuel or driving can at least in principle be varied widely and smoothly enough to touch {\it death}, where equilibrium with the environment is established via the dynamical condition of (global) detailed balance. In other words, death provides a useful theoretical reference or point of contrast for models mimicking biological functioning.  Indeed, one problem with current models of active systems is that there is often no clear physical reference which can be called the equilibrium point from which nonlinear effects such as Life can take off, or perturbation theory at least can meaningfully start.  Usually, there is only the passive limit, such as the asymptotics of large flipping rate for run-and-tumble particles, as the equilibrium point of reference.  While that is formally possible, it is physically odd.  After all, the flip rate is a time-symmetric parameter and does not directly represent any fueling or driving.  Another option is to consider the amplitude of ``other than thermal (white)'' noises as driving strength.  However, that is probably just a rule of thumb and erases all nonMarkovian effects.  Moreover, ``erasing noise'' usually presents a singular limit.  In fact, the point of death is not always  to stop mechanical work by simply putting the amplitude of an active force equal to zero.  Rather, in a biophysical sense, stopping Life is to eliminate inhomogeneities and to stimulate thermalization, and that requires a conceptual unfolding of the underlying machinery.  (Biological) heat engines, while fueled by chemistry, are the natural choice here. This is finally the specific motivation of the present paper, to launch two-temperature models through which important representatives of active processes get a more standard place in nonequilibrium physics.  Away from the reference equilibrium, the condition of local detailed balance ensures proper identification of heat and dissipated power.  We are not the first to propose such unfolding, see \eg \cite{bebon,self,szamel} and the many references therein, but here we insist on the role of the so-called ``hot spots.''\\
 %r\\
 
For modeling a ``hot spot,'' we introduce a model of a flashing vibrational mode in Section \ref{rat}.  It is inspired by the Huxley bridge-mechanism (1957) for attachment and detachment of the myosin motor to the actin substrate. We take that mechanism (in a simplistic form) to be relevant for modeling an activated molecule, first for a harmonic spring in Section \ref{crat} and then, in Section \ref{swit}, in the form of a molecular switch where the order of the energies in a multilevel system gets inverted at random times.   There we show how the activation indeed raises the effective temperature, in the sense of increased uniformization of energy occupation statistics.  The internal temperature (for activation) is found to be the effective temperature in the sense that the occupation statistics of the energy levels starts to resemble an equilibrium system with that much higher or even negative temperature.\\
That naturally leads to an engine model, in Section \ref{eng}, and can be called a form of Brownian ratchet, \cite{brownianratchet, Astumian, Reimann, artibrown, drift, thermalratchet, Brownian_Parrondogame, Brownian_gambling, directmo}, except that we insist (again) on the two-temperature modeling.  The fact that energy transfer and the related possibility to deliver work need not be driven by differences in {\it real} temperature between thermal (equilibrium) baths has been noted before, \eg in \cite{conference, edgarreview, edgarbrowinan, ion, ion2}; it suffices that the ``effective'' temperatures (\eg as calculated from the relative occupations in a multilevel system) between two regions are different.  We thus couple the hot spot to a (cold) flagellum and show that it starts to rotate.  That provides a model for flagellar movement in a medium.  We do not enter there the broader issues of modeling biological engines, but the subject really turns toward run-and-tumble motion.\\
Section \ref{apa} directly deals with models of active particles as two-temperature processes.
Section \ref{rtp} discusses the case of run-and-tumble particles, followed by Section \ref{abp} on active Brownian dynamics.  The two-temperature interpretation now allows to use the condition of local detailed balance, \cite{ldb}.  We observe transitions in the stationary behavior as a function of the system parameters such as the effective temperature, from which robustness of the so-called shape transition, \cite{sol,urtrap}, can be derived, as our modeling allows to extend the phenomenon in the two-temperature setting. From local detailed balance, a thermal characterization becomes possible, \eg, with the proper identification of heat.  Literature that studies dissipation in active processes includes \cite{howfar,nardini}.
For the biological physics of population inversion as visible from calorimetry on active systems, we refer to \cite{activePritha,sim}.  Bringing active particle models under the governing body of local detailed balance is the theoretical contribution to which the present paper wishes to add.

\section{Agitated vibrational modes}\label{rat}

A pioneering model for striated muscle activity was provided by A.F. Huxley in 1957, introducing a cross-bridge mechanism connecting the myosin with the actin filament, \cite{hux}. 
\begin{figure}[H]
    \centering
    \includegraphics[scale = 0.75]{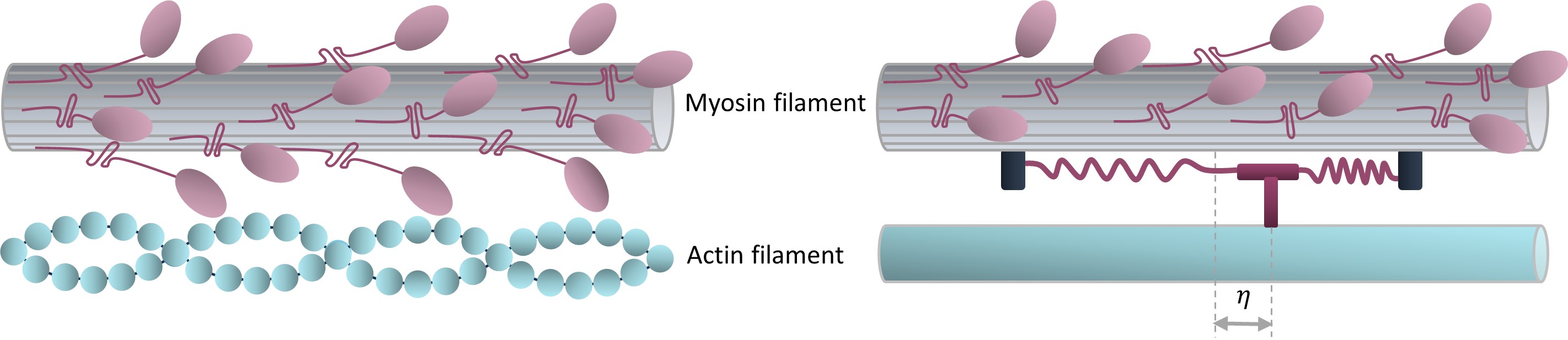}
    \caption{\small{Left: cartoon of the heads of the Myosin molecular motor connecting and disconnecting with the actin filament.  Right:  Huxley bridge-model using  springs, for connecting with the actin filament.  The displacement $\eta$  determines the asymmetric rates of attachment and detachment as in Eq \eqref{od}.}}   \label{huxl}% \label{fila}
\end{figure}
We remain far from studying animal muscles in the present paper but the activity part inspires an elementary model of attachment/detachment dynamics, realized as a switching of a harmonic potential between a convex and a concave shape.  We start with a description in terms of a continuous vibrational mode variable.  A discrete version follows in Section \ref{swit}.

\subsection{Attachment/detachment model}\label{crat}
For describing the value of a vibrational mode in a molecule, we introduce a continuous variable $\eta$, subject to a harmonic potential $\Phi$  with a spring `constant' $\kappa(1\pm w)$  switching values,
\begin{align}\label{hux}
    \Phi(\eta, \sigma)=\frac{\kappa}{2}\, (1+w\, \sigma)\eta^2, \qquad \sigma=\pm 1
\end{align}
The binary variable $\sigma$ is taken to flip at a rate $\alpha \,\exp\{\frac{\beta'}{2}\kappa\,w\sigma\,\eta\} > 0$, for some inverse \emph{internal} temperature $\beta'$ to which we turn later.  
%\begin{figure}[H]
 %   \centering
  %  \includegraphics[scale = 0.85]{huxl.pdf}
   % \caption{\small{Harmonic potential for attachment/detachment dynamics, where in \eqref{hux}$ \kappa=0.5, \omega=2$.}}
    %\label{huxl}
%\end{figure}
We take $\kappa>0$, but the parameter $w$ can be negative and even have $|w|>1$ to allow higher activation.  The vibration mode is then subject to the dynamics %The vibrational energy is still $\kappa\,\eta^2/2$, but the dynamics is
\begin{eqnarray}\label{od}
\dot{\eta}_t &=&  - \kappa\,(1+w\,\sigma_t)\,\eta_t + \sqrt{2 T}\, \xi_t\notag\\
 \sigma_t \longrightarrow - \sigma_t&&\text{  at rate }\quad \alpha \,\exp\{\frac{\beta'}{2}\kappa\,w\sigma\,\eta\}
\end{eqnarray}
As for an overdamped dynamics in a background at temperature $T = \beta^{-1}$, we take $\xi_t$ standard white noise.   We view \eqref{od} as an attachment/detachment model Fig.~\ref{huxl} with the switching rate depending on the position $\eta$ in an asymmetric way as is the case and is essential in the original Huxley model, \cite{hux,herz}.\\

The fact that the \emph{internal} $\beta'$ does not (necessarily) equal the ambient $\beta$ means to indicate the resistance of the molecule to rapid thermalization. The word `internal' refers to the internal fueling or agitation by biochemical processes that fall outside the present description.
To shed light on the role of that `internal' temperature, we next estimate an \emph{effective} temperature from the spreading $\langle \eta^2\rangle^s$ in the stationary distribution (if it exists), with stationary expectation indicated by $\langle\cdot\rangle^s$.  That appears to be a good choice from the point of view of response theory as well, \cite{yael}.\\

We first take the case where $\beta'=0$ (having then the switching $\sigma\rightarrow -\sigma$ at constant rate $\alpha$).  Solving the kinetic equations leads to the following result for the steady condition, \cite{activePritha}: \\
When $|w| > 1$, there is a critical persistence
\[
z_c = \frac 1{w^2-1}
\]
in the sense that there is no stationary distribution for $\kappa \geq \alpha\,z_c$.  For $|w| \leq 1$, the second moment in the stationary distribution  equals
\[
	\langle \eta^2\rangle^s = \frac{\alpha + \kappa}{\alpha + \kappa(1- w^2)} \frac{T}{\kappa} = \frac{T_\text{eff}}{\kappa}
\]
We see the usual equilibrium for $w=0: \langle \eta^2\rangle^\text{eq} = T/\kappa$.   When $|w| < 1$, the effective temperature $T_\text{eff}$ is slightly larger than $T$.  When $|w|> 1$, the stationary second moment $\langle \eta^2\rangle^s$ diverges for $\kappa/\alpha \uparrow z_c$, which means that the effective temperature reaches infinite as the flip rate $\alpha\downarrow \alpha_c = \kappa\, (w^2-1)$.  Then indeed, the system increasingly occupies very high energy states in that single vibrational mode.  It has become a ``hot spot'' indeed.\\
The situation is similar for small $\beta'$, but the formul{\ae} become more complicated.  To be more explicit, we turn to the analogue in an exactly solvable case of discrete (equally spaced) energy levels in the limit $\alpha\uparrow \infty$.

\subsection{Multilevel switch}\label{swit}
As a discrete example of an agitated molecule, we consider a switch.
Molecular switches are proteins that flip between two architectures causing either an {\it on}- or {\it off}-state for intracellular signaling. The {\it on} is an active state, typically initiated from an external stimulus.  The change in energies may be provoked by the (sudden) entrance of an ion, a proton, or even a photon that modifies the molecular configuration or the harvesting of energy on even smaller (nano)scales.  Such processes happen on the level of individual molecules.\\

\begin{figure}[H]
    \centering
    \includegraphics[scale = 0.67]{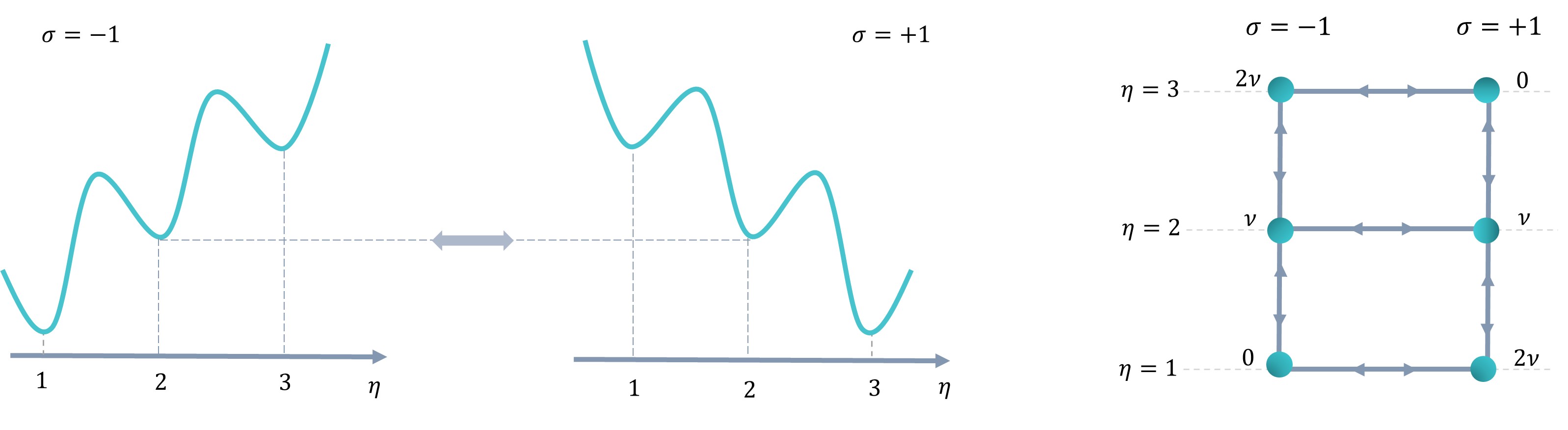}
    \caption{\small{Modeling a three-level switch ($n=3$). Left: switching the energy landscape. Right: graph of transitions with an indication of the energies \eqref{ener}.  Vertical transitions happen at ambient temperature; horizontal transitions at `internal' temperature.}}
    \label{tobemade}
\end{figure}

One type of modeling proceeds via specifying a multilevel system where, \eg at random times, energy levels are shifted.   Here we follow a discrete version of the Huxley harmonic modeling in the previous section. We imagine an energy labeled by a coordinate  $\eta=1,\ldots,n$, which is mirrored by flipping a variable $\sigma = \pm 1$ indicating the {\it on}- or {\it off}-state.      More precisely, we consider energies 
\begin{equation}\label{ener}
E(\eta,\sigma) = (n-1+(n-2\eta+1)\sigma)\,\frac{\nu}{2}
\end{equation}
separated by the same energy unit $\nu>0$ at fixed $\sigma$. We refer to Fig.\ref{tobemade} for describing the situation for $n=3$.  
Transitions between energy levels are described by a Master equation for transition rates, 
\begin{align}
\text{ if } \sigma=-1:\, & \; \eta\rightarrow \eta+1 \text{ at rate } \frac {1}{\tau}e^{-\frac{\beta\nu}{2}} \text{ and } \eta+1 \rightarrow \eta \text{ at rate } \frac {1}{\tau}e^{\frac{\beta\nu}{2}}\label{bar}\\
\text{ if } \sigma=+1:\, & \;\eta\rightarrow \eta+1 \text{ at rate } \frac{ 1}{\tau} e^{ \frac{\beta\nu}{2}} \text{ and } \eta+1 \rightarrow \eta \text{ at rate } \frac {1}{\tau}e^{-\frac{\beta\nu}{2}}
    \notag\\
\sigma \rightarrow -\sigma &\text{ at rate }\quad \alpha\,e^{\frac{\beta'}{2}(E(\eta,\sigma)-E(\eta,-\sigma))} = \alpha\,e^{\beta'\,\nu\,(n-2\eta+1)\,\sigma}\label{mode}
%   k_\eta(\sigma)&= \alpha  \label{ladder} %\,e^{\frac{\beta(1-\ve)}{2}(E(\eta,\sigma)-E(\eta,-\sigma))}
\end{align}
for ambient inverse temperature $\beta$, internal inverse temperature $\beta'$, and where $\alpha\tau>0$ indicates the switching amplitude.  That Markov jump dynamics $(\eta_t,\sigma_t)$ makes the analogue of \eqref{od}.  Of course, at $\eta=1$ and $\eta=n$, the system state can only move up or down the ladder, or switch stile.\\
 
\begin{figure}[H]
     \centering
      \begin{subfigure}{0.49\textwidth}
         \centering
         \def\svgwidth{0.8\linewidth}        
        \includegraphics[scale = 0.76]{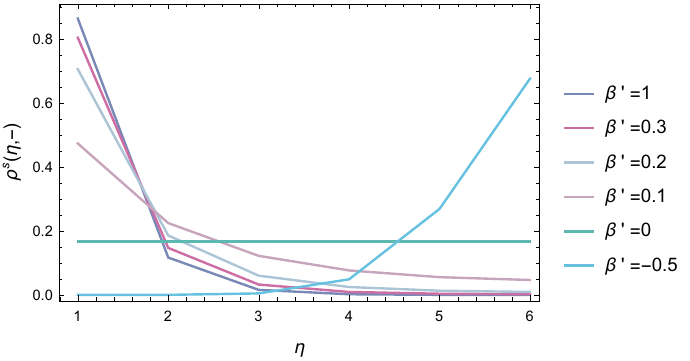}
     \end{subfigure}
     \hfill
     \begin{subfigure}{0.49\textwidth}
         \centering
         \def\svgwidth{0.8\linewidth}        
   \includegraphics[scale = 0.76]{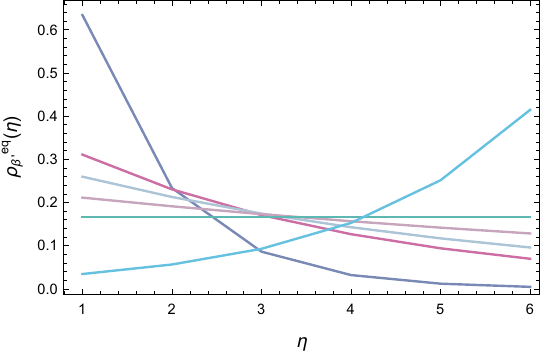}
     \end{subfigure}
\caption{The internal temperature is an effective temperature.  \small{Left: the stationary occupation $\rho^s(\eta,
\sigma=-1), \eta=1,\ldots, 6$ in the fast-switching limit $\alpha\uparrow\infty$ for different $\beta'$ at $\tau=1,\nu=1,\beta=2$ for $n=11$.  The exact solution was obtained thanks to private discussions with Hidde Van Wiechen.  Right: the equilibrium profiles $\rho^\text{eq}_{\beta'}$ computed at the inverse internal temperature $\beta'$ are similar to the profiles on the left.}} \label{hidden}
\end{figure}

We are interested in the stationary distribution $\rho^s_{\beta,\beta',\alpha}$ and its dependence on $(\beta,\beta',\alpha)$.
Detailed balance holds true for $\beta' =\beta$ (at all $\alpha>0$), with energy
\eqref{ener}.  The corresponding equilibrium distribution $\rho^s_{\beta,\beta,\alpha} = \rho^\text{eq}_\beta$ is given by the Boltzmann--Gibbs weight and is exponentially decaying from $\eta=1$ to $\eta \simeq n/2$ when $\sigma=-1$.  Whatever $\beta$, when $\beta'=0$, there is a (purely) random flipping between the two energy ladder stiles. Then, the switch occurs at rate $\alpha$ independently of the molecular state $\eta$, and no feedback occurs.  
In general, the $\alpha\uparrow \infty$-limit of the stationary distribution $\rho^s_{\beta,\beta',\alpha}$ is exactly solvable and has a nontrivial profile  $\rho^s(\eta) = \lim_\alpha\sum_{\sigma =\pm 1}\rho^s_{\beta,\beta',\alpha}(\eta,\sigma)$.  From it, we can also derive the limiting energy-level occupation $\rho^s(\eta,-1) $. We see it for $n=11$  in the left plot of \fig\ref{hidden}.
 
As we move from $\beta'=\beta$ to $\beta' =0$, always for large $\alpha$, the profile $\rho^s(\eta)$ over $\eta$ becomes more and more uniform.  In other words, the nonequilibrium driving adds weight to the levels that are less occupied in equilibrium, realizing effectively an occupation statistics corresponding to a higher temperature (infinite for $\beta'=0$ ). It is therefore again a matter of consistency to call $\beta'^{-1}$ not only the \emph{internal} (switching) temperature but to call it an \emph{effective} temperature as well. To witness, in Fig.\ref{hidden} we have added the equilibrium profiles $\rho^\text{eq}_{\beta'}$;  they indeed resemble the profile of $\rho^s_{\beta,\beta',\alpha\uparrow\infty}$ for $\beta'\ll \beta$. In the large switching regime and for smaller ambient temperatures, the internal temperature has become effective, making the molecule a ``hot spot.''\\

\begin{figure}[H]
     \centering
      \begin{subfigure}{0.49\textwidth}
         \centering
         \def\svgwidth{0.8\linewidth}        
        \includegraphics[scale = 0.74]{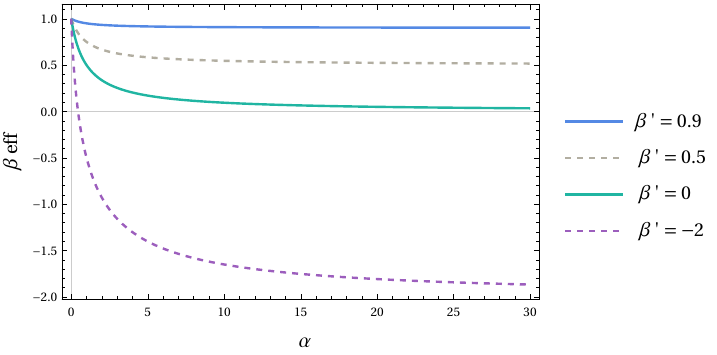}
     \end{subfigure}
     \hfill
     \begin{subfigure}{0.49\textwidth}
         \centering
         \def\svgwidth{0.8\linewidth}        
   \includegraphics[scale = 0.74]{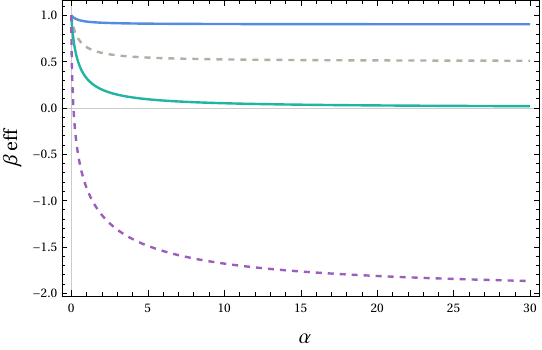}
     \end{subfigure}
\caption{\small{The effective inverse temperature \eqref{eft} for $\beta=\nu=1$, as function of $\alpha$. Left: $n=2$; Right: $n=3$, where $\beta_\text{eff} = \frac 1{2 \nu} \,\log \frac{\rho(1,-)}{\rho(3,-)}$ (courtesy of Drieke Keuppens and Luca Dehennin).}}   \label{ec}
\end{figure}

 Typically, $n$ can be considered small.   For $n=2$ we are dealing with a switch for a two-level system (or classical qubit) where at random times the ground state and the excited state get exchanged, \cite{jchemphys,sim}. To be specific about effective temperature  we define 
 \begin{equation}\label{eft}
%k_B\,T_\text{eff} = \nu \,\log \frac{\rho(2,-)}{\rho(1,-)}
\beta_\text{eff} = (k_B\,T_\text{eff})^{-1} = \frac 1{\nu} \,\log \frac{\rho(1,-)}{\rho(2,-)}
\end{equation}
measuring the homogenization of energy occupation; it is plotted in the left Fig.~\ref{ec}.  Similar definitions are possible for larger $n$ as is \eg done for obtaining the right plot of Fig.\ref{ec}.
Lowering $\beta'\geq 0$ makes the stationary occupations uniform, after which, for $\beta'<0$ and at least for large enough $\alpha$, leads to population inversion. It implies that the effective temperature \eqref{eft} increases from $T_\text{eff} = T$ at $\beta'=\beta$, to $T_\text{eff} = \infty$ at $\beta'=0$, and then becomes negative for $\beta' < 0$ (for large enough switching rate $\alpha)$; see Fig.~\ref{ec} qualifying the model as describing a hot spot for $\beta'\simeq 0$.

To a good approximation $\beta_\text{eff} = \beta'$ in the limit $\alpha\uparrow\infty$.  From the facts above, we conclude (again) that the internal switching temperature is an effective temperature for representing energy-level occupations.  The `effective' has no other meaning than in the discussion above, and in particular is not directly connected to restoration of a fluctuation--dissipation relation, \cite{cug}.
%Leticia F. Cugliandolo, The effective temperature. J. Phys. A (Review Section) 44, 483001 (2011).

\section{Engine model}\label{eng}
The next step is to get work done.  We couple therefore the agitated vibrational mode from the previous section to a work variable, an angle variable $y\in S^1$ on the circle.  We think of that $y$ as the rotation angle of a bacterial flagellum.  To be specific, proceeding first with the discrete setting of Section \ref{swit}, we consider now an energy function
\[
V(y,\eta,\sigma) = U(\eta,\sigma,y) + E(\eta,\sigma)
\]
where $E$ comes from \eqref{ener}, and the coupling $U(\eta,\sigma,y) = \lambda \,\cos(y - \eta)$ when $\sigma=-1$, and zero otherwise. That choice is rather arbitrary as we do not discuss the action of the molecular motors.  The angular dynamics for $y_t$ then supposedly follows the overdamped equation of motion
\begin{equation}\label{fla}
\dot{y}_t = \lambda\,\sin(y_t -\eta_t)\,\frac{1-\sigma_t}{2}
\end{equation}
while the $(\eta_t,\sigma_t)$ is a jump process with transition rates that depend on the angle $y$,
\begin{align}
 & \; \eta\rightarrow \eta\pm 1 \text{ at rate } \quad\frac {1}{\tau}e^{-\frac{\beta}{2}[V(y,\eta\pm 1,\sigma)-V(y,\eta,\sigma)]} \label{baro}\\
& \; \sigma \rightarrow -\sigma \text{ at rate }\quad \alpha\,e^{\frac{\beta'}{2}(V(y,\eta,\sigma)-V(y,\eta,-\sigma))} \label{modeo}
%   k_\eta(\sigma)&= \alpha  \label{ladder} %\,e^{\frac{\beta(1-\ve)}{2}(E(\eta,\sigma)-E(\eta,-\sigma))}
\end{align}
For $\lambda=0$, those formul{\ae} reduce to \eqref{bar}--\eqref{mode}.\\
  Obviously, from \eqref{fla}, there is no angular motion when $\sigma_t=1$, and at those times the particular flagellum is not rotating and, depending on the activity of other flagella, the bacteria will run in another direction.  That is the typical run-and-tumble motion we are after.\\
To check there is a rotational force, we take the stationary expectation over $(\eta,\sigma)$,
\begin{equation}\label{rotf}
f(y) = \lambda\,\sum_{\eta=1}^n \sin(y -\eta)\,\rho^s(\eta,\sigma=-1)
\end{equation}
It is the dominating force in the case of time-scale separation between the fast 'active' variable $\eta$ and the slower angle $y$; see \cite{nong, bram2}. \\

\begin{figure}[h]
    \centering
    \includegraphics[scale=0.9]{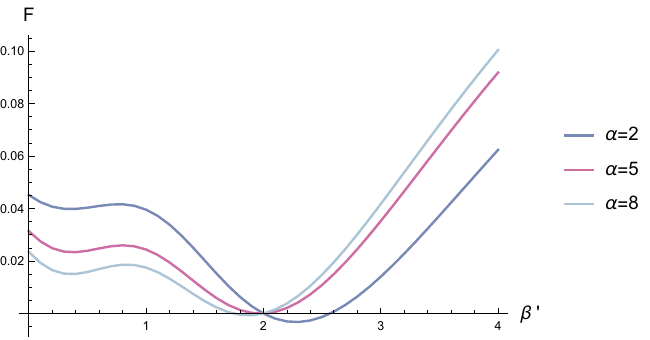}
    \caption{\small{Rotational force $F = \oint f(y) \id y $,  as a function of the internal $\beta'$ for different $\alpha$, where $n=3$, $\lambda=-1, \beta=2, \nu=1 $.}}
    \label{force}
\end{figure}

Fig.~\ref{force} shows that rotational force $F= \oint f(y)\,\id y$; it provides the flagellar movement of a bacteria, and the induced motility. Note that it vanishes at detailed balance when $\beta=\beta'$, but there may be other temperatures where the rotational component vanishes.\\

It requires little to have other versions, \eg, that the rotation would be inverted when $\sigma$ flips, or that the rate of switching would be asymmetric to allow for different time scales between different types of running.\\

There also exists a version by taking the
continuous mode of Section \ref{crat}.  We  couple the (now) real variable $\eta$ with  angle variable $y\in S^1$, as 
\begin{eqnarray}
\dot{\eta}_t &=&  - \partial_\eta U(y_t-\eta_t)- \kappa\,(1+w\,\sigma_t)\,\eta_t + \sqrt{2 T}\, \xi_t^{(\eta)}\notag\\
m\ddot{y} + \gamma \dot y_t &=& - \partial_y U(y_t-\eta_t) + F + \sqrt{2m\gamma T}\, \xi_t^{(y)}\label{flag}
\end{eqnarray}
with periodic potential $U$ and periodic force $F$.    %; e.g. $U(y) = \lambda y^2/2$ for a harmonic coupling.\\
The same construction allows longer chains and/or overdamped probes $y_t$, directly defined on $\bbR^d$.  As the $\eta$-dynamics violates detailed balance for $\beta'\neq \beta$, the angle variable $y_t$ will start to rotate while carrying the constant load $F$.
As above we get flagellar locomotion, as the rotation can give rise to spatial translation in the medium.

\section{Active particles}\label{apa}
The previous section realizes active models of self-locomotion in a two-temperature setting, with one high effective temperature obtained from the dynamics of an internal variable (hot spot). In what follows we take over the main structure of that dynamics but skip the flagellar variable.  As a shortcut, we directly couple the activity to the position of the bacteria.

\subsection{Run-and-tumble particles}\label{rtp}
\begin{figure}[h!]
    \centering
    \includegraphics[scale=0.45]{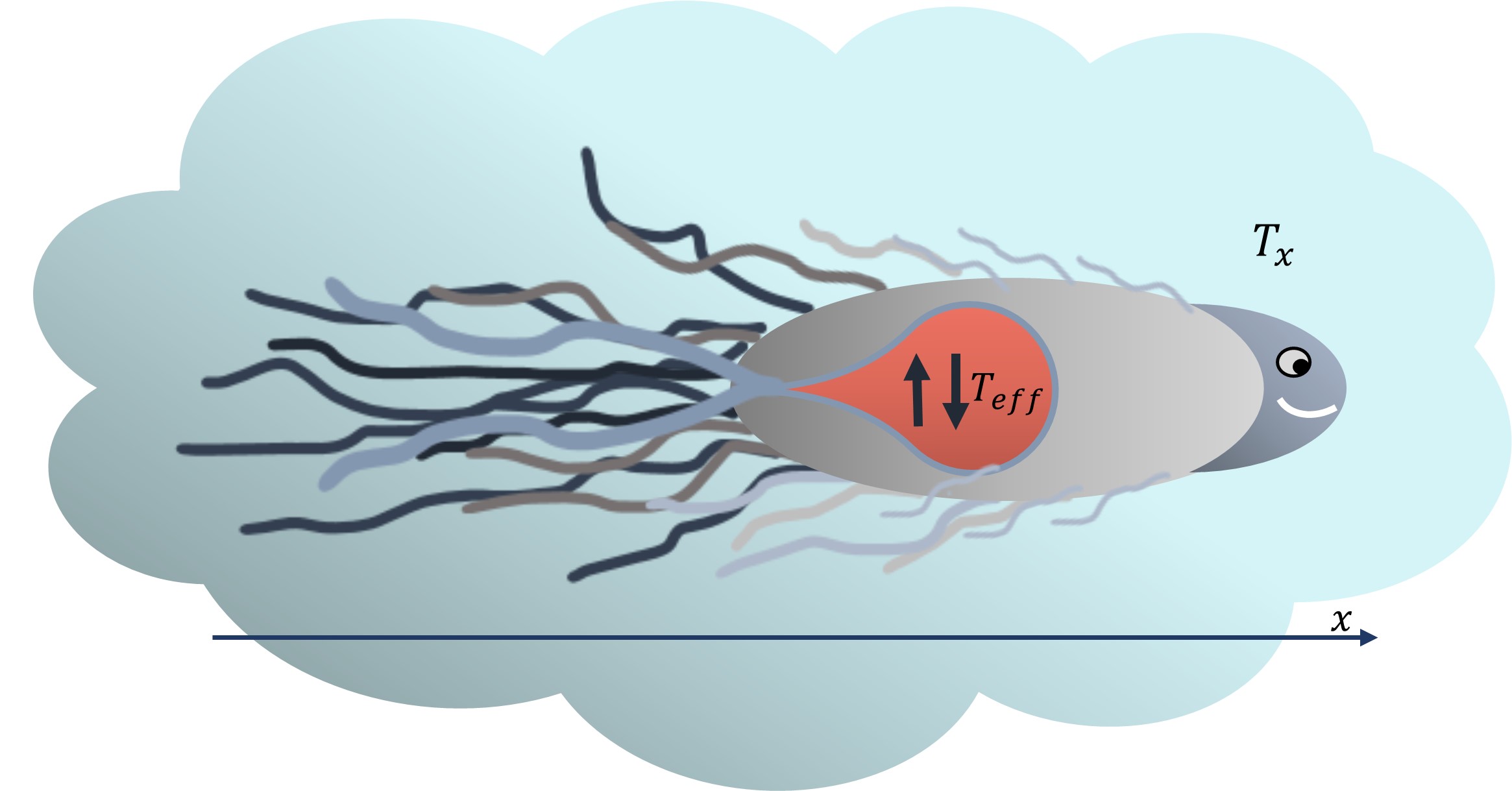}
    \caption{Cartoon of an active particle, for zooming in on the ``hot spot'' at effective temperature $T_\text{eff}$, coupled to the locomotion system at ambient temperature $T=T_x$, with $x$ the position of the overdamped swimmer.  The biological interpretation of the ``hot spots'' will certainly differ per organism.  For {\it E. Coli} as modeled via RTPs in the present section, we have in mind motility proteins (the rotary engine) with power provided by a proton flow, while for Eukaryotes, it is tempting to think about mitochondria.}
    \label{ecc}
\end{figure}

For simplicity we consider position $x\in \bbR$ subject to the dynamics,
\begin{equation}\label{con}
\dot x _t= v\sigma_t - U'(x) + \sqrt{2\gamma\, T}\,\xi_t
\end{equation}
 The potential $U$ is spatially confining around the origin. The $\xi_t$ is standard white noise with ambient temperature $\beta^{-1} =k_BT$ giving it strength.  The hot spot dynamics is represented by
$\sigma_t = \pm 1$, which flips sign
\begin{equation}\label{tr}
\sigma\rightarrow -\sigma \;\text{ with rate } \;\; \alpha\,\exp{-\beta_\text{eff} \,v\, \sigma\, x}
\end{equation}
depending on the position $x$ of the walker.  For $\beta_\text{eff}=0$, the $\sigma_t$ represents standard dichotomous noise, and then \eqref{con} (often with $T=0$) describes a one-dimensional run-and-tumble motion with tumbling rate $\alpha$.  Fig.~\ref{ecc} gives an impression of the situation we wish to imagine more generally.  \\

At first sight, there is some problem with the tumbling rates \eqref{tr}, as they depend on the location of the spatial origin.  That is however the same origin as already specified by the confining potential.  Moreover, from bacterial chemotaxis, spatial gradients in flipping rate are not uncommon.  Furthermore, for $\beta_\text{eff} \simeq 0$ (high internal temperature) that dependence becomes very weak, and for the locomotion part \eqref{con}, the force $v\,\sigma_t$ is fully ignorant about the position. The main thing however is to see it as a continuation from the models of the previous section, skipping the flagellar motion, taking the dichotomous noise now acting directly on the translational motion.\\
The great advantage of the parametrization in \eqref{con}--\eqref{tr} is that we have a clear reference  $\beta_\text{eff}=\beta$ (keeping $\alpha\neq 0$) where the condition of detailed balance holds.  Fixing $\alpha>0$, we can now interpolate between the purely random flipping ($\beta_\text{eff}=0$) and the equilibrium dynamics ($\beta_\text{eff}=\beta$) via that temperature-driven analog.  At the same time, it brings the RTP--modeling within the realm of local detailed balance \cite{ldb} from where thermal features are more easily interpreted.  
There is a contact between (effectively) two  thermal baths, one at inverse (ambient) temperature $\beta$, and one at inverse (internal but effective) temperature $\beta_\text{eff}$.  
The difference  $\beta-\beta_\text{eff}$ in inverse temperatures provides the driving, much as in a traditional heat engine.  In that way, we interpret the motion of RTPs as originating from a two-temperature model, from which the usual thermodynamic interpretation of heat flux holds.

\subsubsection{Heat}
The parametrization in the two-temperature process \eqref{con}--\eqref{tr} easily identifies the driving and permits a standard discussion of thermal properties. Since the system is open (to two baths) the energy $E(x,\sigma) = -v\sigma x + U(x)$ changes and it does so in two ways: %from \eqref{con},
\[
\frac{\id}{\id t}\langle E(x_t,\sigma_t)\rangle = \big\langle (L_x + L_\sigma)E\,(x_t,\sigma_t)\big\rangle
\]
for
\begin{eqnarray}\label{lx}
L_xE(x,\sigma) &=& (-v\sigma + U'(x))\,[v\sigma - U'(x)] + T\,U''(x)\\
L_\sigma E(x,\sigma) &=&  2v\sigma\,x\,\alpha\,\exp{\frac{\beta_\text{eff}}{2} [E(x,\sigma)-E(x,-\sigma)]}\nonumber
\end{eqnarray}
Therefore, upon specifying the instantaneous variables $(x,\sigma)$,
\begin{align}\label{energ}
&\frac{\id}{\id t}\langle E(x_t,\sigma_t)\,|\,(x(0)=x,\sigma(0)=\sigma\rangle_{|_{t=0}} + \dot{q}_x(x,\sigma) + \dot{q}_\sigma(x,\sigma) = 0\\
&\text{ for } \qquad
\dot{q}_x(x,\sigma) = [v\sigma - U'(x)]^2 - T\,U''(x)\\
& \hspace{0.83cm} \qquad \dot{q}_\sigma(x,\sigma) = -2v\sigma\,x\,\alpha\,\exp{-\beta_\text{eff} v\, \sigma\, x}
\end{align}
where $\dot{q}_x(x,\sigma)$ %= [v\sigma - U'(x)]^2 - T\,U''(x)$ 
equals the expected heat flux to the ambient bath at inverse temperature $\beta$, and 
$\dot{q}_\sigma(x,\sigma) $
%=-2v\sigma\,x\,\alpha\,\exp{-\beta (1-\ve) v\, \sigma\, x}$ 
is the expected heat flux to the internal bath at inverse temperature $\beta_\text{eff}$.  Of course, in the steady distribution, their sum $\langle \dot{q}_x(x,\sigma) + \dot{q}_\sigma(x,\sigma)\rangle^s$ (expected value in the stationary distribution) equals zero, and there is a constant flow of energy between the two heat baths.  From the above, as is common practise now by using standard methods such as outlined in \cite{jmp, heatcon, jchemphys}, various thermal properties can be derived such as variable and mean entropy production, heat capacities, and delivered work in heat conduction networks.

\subsubsection{Shape transition}
 
We are interested in observing the shape transition between an edgy and a confined regime as a function of the effective temperature.\\

The process satisfies detailed balance for $\beta_\text{eff}=\beta$ (equal temperatures), with 
 equilibrium density
\begin{equation}\label{eqr}
\rho_\text{eq}(x,\sigma) \propto e^{-\beta E(x,\sigma)}, \qquad E(x,\sigma) = -v\sigma x + U(x)
\end{equation}
at (unique) inverse temperature $\beta$ (independent of $\alpha$, as it should), at least as long as the potential $U$ is sufficiently confining.  For example, for the harmonic potential $U(x)=x^2$, we have
\begin{equation}\label{eqrpot}
\rho_\text{eq}(x,\sigma) \propto e^{-\beta (x-v\sigma/2)^2}
\end{equation}
and it depends on the propulsion speed $v$ and on $\beta$ whether $\rho_\text{eq}(x)= (\rho_\text{eq}(x,+1)+ \rho_\text{eq}(x,-1))/2$ has its maximum at the origin $x=0$ (confined regime)  or at the edges $x=\pm v/2$ (edgy regime).  
        
\begin{figure}[H]
     \centering
     \begin{subfigure}{0.49\textwidth}
         \centering
         \def\svgwidth{0.75\linewidth}        
        \includegraphics[scale = 0.85]{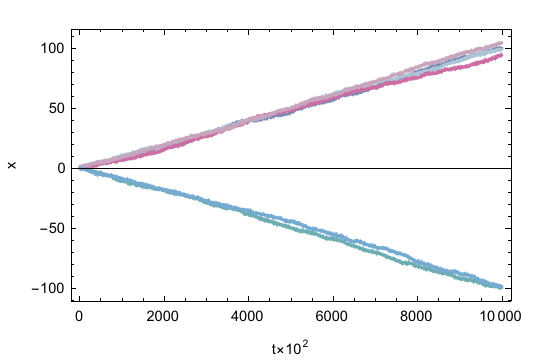}
         \caption{\small{For $n=6$ at $\beta_\text{eff}=0.5\beta$}}
     \end{subfigure}
     \hfill
     \begin{subfigure}{0.49\textwidth}
         \centering
         \def\svgwidth{0.8\linewidth}        
   \includegraphics[scale = 0.85]{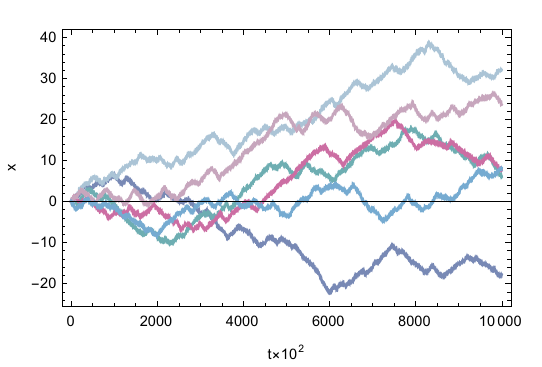}
         \caption{\small{For $n=6$ at $\beta_\text{eff}=0$}}
     \end{subfigure}
          \hfill
     \begin{subfigure}{0.49\textwidth}
         \centering
         \def\svgwidth{0.8\linewidth}        
   \includegraphics[scale = 0.85]{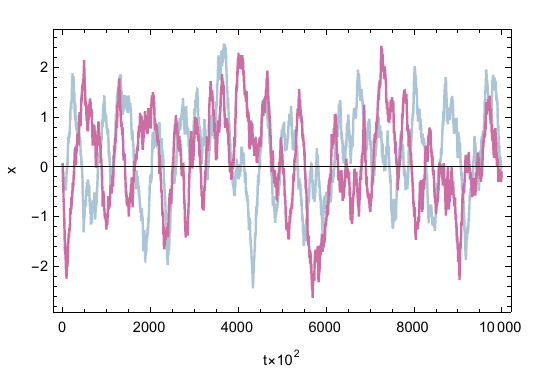}
         \caption{\small{For $n=2$ at $\beta_\text{eff}= -0.1\beta$}}
     \end{subfigure}
    \caption{\small{Trajectories of $n$ particles under  free low-temperature locomotion  \eqref{con}--\eqref{tr} with $U=0$, $v=1, \beta=10,\alpha=0.5$, for different $\beta_\text{eff}$, from low(a) to high(b)--(c) internal temperatures.}}\label{stat}
        \end{figure}

We are interested mostly in large $\beta$  (small thermal noise for locomotion).\\
Suppose $0<\beta_\text{eff}\leq\beta$: the flipping rate  $\alpha\,\exp{-\beta_\text{eff}\, v\, \sigma\, x}$ is going to be small when $\sigma\,x > 0$ and $\beta_\text{eff}$ is large, and hence, say for $\sigma=1$, the particle will continue to large $x>0$ without flipping (flipping rate decreasing ever more).  For example, for $\sigma\,x\sim v$, the particle is expected to get trapped for a long time, especially at large propulsion speed $v$, and $\alpha$ needs to be truly large to avoid that.  For example, it remains true at $T=0$, from \eqref{con}, that $\dot x_t=0$ when $x=\sigma\,v/\kappa$ for the harmonic potential $U(x) =\kappa x^2/2$.   Without confinement ($U=0$) and with little thermal noise ($\beta=10$), the trajectories look as plotted in Fig.~\ref{stat}(a).  That is giving rise, as for $\beta_\text{eff}=0$, to edge states; compare Fig.~\ref{stat}(a) with Fig.~\ref{stat}(b) which represents trajectories in the standard RTP-process for which $T=0, \beta_\text{eff}=0$.\\

\begin{figure}[H]
 \centering
        \begin{subfigure}{0.45\textwidth}
         \centering
         \def\svgwidth{0.45\linewidth}        
   \includegraphics[scale = 0.65]{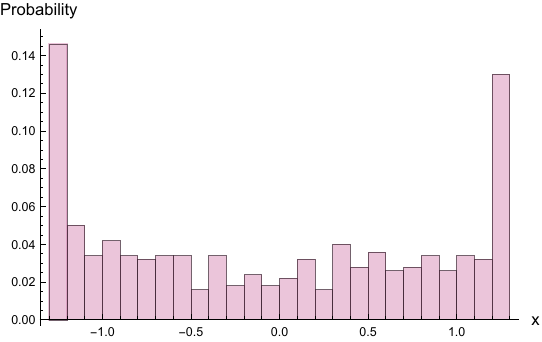}
         \caption{\small{Edge state, visible in the stationary distribution at $\alpha=2 \quad \beta_\text{eff}=0.1$}}
     \end{subfigure}
         \hfill
     \begin{subfigure}{0.45\textwidth}
         \centering
         \def\svgwidth{0.45\linewidth}        
   \includegraphics[scale = 0.65]{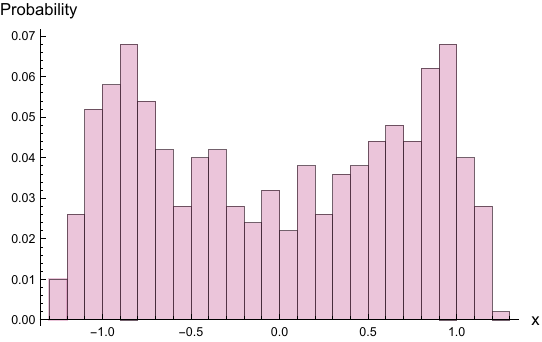}
         \caption{\small{Edge state, visible in the stationary distribution at $\alpha=30\quad \beta_\text{eff}=0.4$}}
     \end{subfigure}
          \hfill
            \begin{subfigure}{0.45\textwidth}
         \centering
         \def\svgwidth{0.45\linewidth}        
        \includegraphics[scale = 0.65]{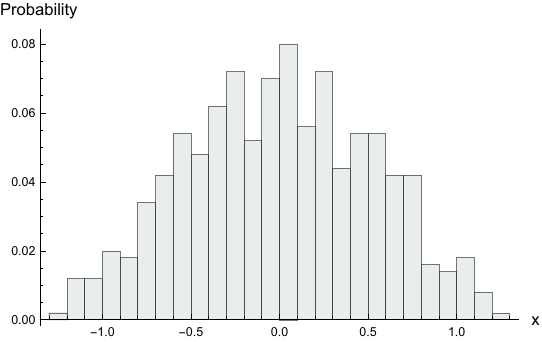}
         \caption{\small{Confinement at $\alpha=10,\quad  \beta_\text{eff}=0.1$}}
     \end{subfigure}
     \caption{\small{Stationary density of the two-temperature RTP-process  at $v=3, T=0$ with confining potential $U=x^2$. For large enough $\alpha$, depending on $\beta_\text{eff}$ and $v$, the two peaks at the edges merge and we enter the confined regime with a single peak at the origin. }}\label{epalpha}
        \end{figure}

\begin{figure}[H]
            \centering
             \includegraphics[scale = 0.8]{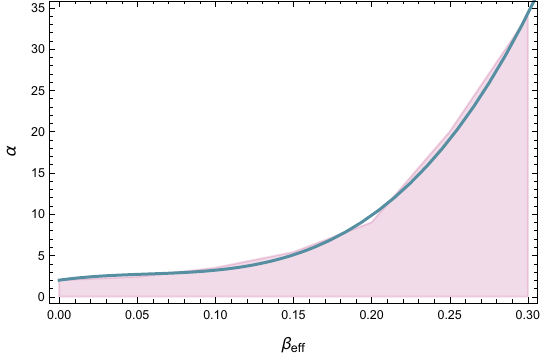}
           \caption{\small{Shape transition in the ($\beta_\text{eff}, \alpha$)-plane for harmonic potential $U(x)=x^2$ at $T=0$ and with propulsion speed $v=3$. The colored section represents the edgy regime. For high enough $\alpha$ starts the confined regime, with a sharply increasing transition line, fitting the transition as observed numerically by coloring 40 points  depending on the  concavity (colored) {\it versus} convexity of the simulated stationary density at the origin; see also \fig\ref{epalpha}. }}
            \label{trans}
        \end{figure}

Suppose next that $\beta_\text{eff}<0$.  The hot spot has reached a negative temperature (indicating population inversion for a multilevel system). The flipping rate  $\alpha\,\exp{-\beta_\text{eff}\, v\, \sigma\, x}$ is going to be small when $\sigma\,x < 0$, and large when  $\sigma\,x > 0$.  Hence, if initially the particle is pushed with $\sigma =1$ to the positive $x>0$, it will flip a lot and its motion remains bounded.  Similarly, when $\sigma=-1$ and pushed to negative $x<0$,  it will flip a lot, and return to the origin.  That implies that for $\beta_\text{eff}<0$ the particles remain confined around the origin even for relatively small $\alpha$, and even when $U=0$ (no confining potential) at small $T$.  The typical trajectories (for $U=0, \beta=10,\, \beta_\text{eff}<0$) are shown in Fig.~\ref{stat}(c).  For $\beta_\text{eff}\simeq\beta$, we need a large flipping rate to see the (passive) confined regime, as illustrated by comparing  Fig.~\ref{epalpha}(a) and (b) with Fig.~\ref{epalpha}(c). See Fig.~\ref{trans} for the boundary of the two regimes.  Note that it includes high $\beta_\text{eff}$, entering the close-t0-equilibrium or McLennan regime. Of course, there, the flip rate $\alpha$ must be very high to get edge states. \\
For $\beta_\text{eff}> \beta$, we would have that the hidden bath has a lower temperature than the environment, which would make changes in configuration less efficient (as if the biological process must warm up the hidden degrees of freedom, instead of using their energy).\\

%\cite{simrtp}???

\subsection{Active Brownian motion}\label{abp}
\begin{figure}[H]
     \centering
      \begin{subfigure}{0.49\textwidth}
         \centering
         \def\svgwidth{0.75\linewidth}        
        \includegraphics[scale = 0.85]{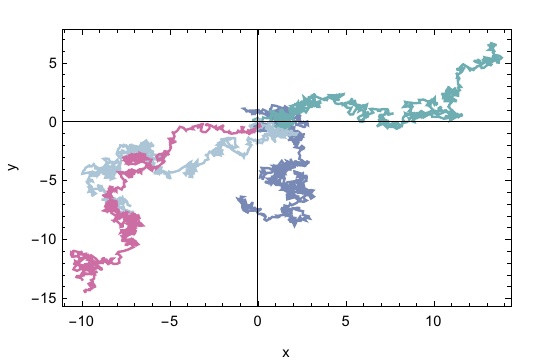}
        % \caption{$a=1, \ve=0.1, \beta=2,$}
     \end{subfigure}
     \hfill
     \begin{subfigure}{0.49\textwidth}
         \centering
         \def\svgwidth{0.8\linewidth}        
   \includegraphics[scale = 0.85]{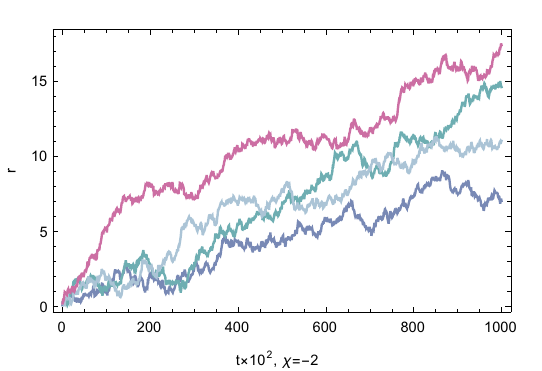}
         %\caption{\small{$a=1, \ve=0.1,  \alpha=1$}}
     \end{subfigure}
\caption{\small{Active Brownian motion for four particles with the dynamics \eqref{abm} at  $\chi<0$, when   $v=R=D=1$ and $U=0$ . Left: trajectories when the particles start from $r(0)=0$. Right: the distance of the particle positions from the origin. }} \label{2dn}
\end{figure}

\begin{figure}[H]
     \centering
      \begin{subfigure}{0.49\textwidth}
         \centering
         \def\svgwidth{0.75\linewidth}        
        \includegraphics[scale = 0.85]{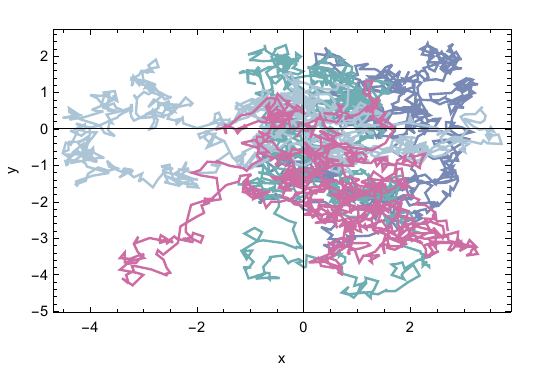}
        % \caption{$v=1, V=0, R=D=1, x(0)=0$}
     \end{subfigure}
     \hfill
     \begin{subfigure}{0.49\textwidth}
         \centering
         \def\svgwidth{0.8\linewidth}        
   \includegraphics[scale = 0.85]{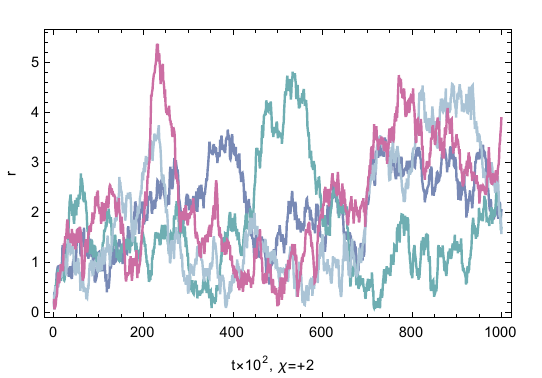}
         %\caption{\small{$v=1, V=0, R=D=1, x(0)=0$}}
     \end{subfigure}
\caption{\small{Active Brownian motion  for four particles  for the dynamics \eqref{abm} with $\chi>0$, when   $v=R=D=1$ and $U=0$ . Left: trajectories when the particles start from $r(0)=0$.  Right: the distance of the particle positions to the origin.}} \label{2dp}
\end{figure}

We move to two dimensions with position $\vec r$ for a point particle, satisfying the equation of motion in Cartesian coordinates $\vec r = (x,y)$,
\begin{eqnarray}\label{abm}
\dot{\vec r}(t) &=& v\,\hat{n}(t) - \mu\,\vec{\nabla} U(\vec r(t)) + \sqrt{2D}\,\vec{\xi}(t)\notag\\
\dot{\theta} &=& \chi[v\,x\,\sin\theta -  v\,y\,\cos \theta]+ \sqrt{2R}\,\zeta(t)
\end{eqnarray}
where the unit vector $\hat{n} = (\cos \theta,\sin \theta)$ replaces the dichotomous noise in \eqref{con}. As before, $\vec{\xi}$ is standard white noise on the two-dimensional locomotion of the particle, while $\zeta(t)$ is rotational white noise, constantly redirecting the particle.  The external potential $U$ is confining and $\mu >0$ is a mobility coefficient. The constant $\chi$ can be taken as positive or negative. Note that we still deal with a point particle and the $\theta-$dynamics is indicating the direction of locomotion.\\
We basically repeat the same remarks and analysis of the previous section on RTPs. The $(x,y)-$dependence in  \eqref{abm} is not visible in the locomotion (first equation for $\dot{\vec r}$ ) and refers to a spatial frame where the origin is determined by the confining potential $U$.\\
The ambient temperature $T$ and the internal effective temperature $T_\text{eff}$
follow from
\[
k_BT = \frac{D}{\mu},\qquad k_BT_\text{eff} = \frac{R}{\chi\,\mu} %= \frac{v}{\ell\chi}
\]
For $\chi=0$ we get the standard active Brownian process, which, in the present setup, refers to infinite effective temperature $T_\text{eff}=\infty$ (as it was for standard run-and-tumble particles as well); see {\it e.g.}, \cite{ur2d}.\\
In our setup and with a confining potential $U$, we imagine an energy function 
\[
E(x,y;\theta) = -\frac{v}{\mu}x\,\cos\theta - \frac{v}{\mu}\,y\,\sin\theta + U(x,y)
\]
coupling the hot spot (with coordinate $\theta$) with the spatial coordinates $(x,y)$.  There is {\it a priori} no problem with taking $T_\text{eff}<0$ via $\chi < 0$ as that temperature is effective and may refer to population inversion.\\
There are again two equilibrium cases. The persistence length is $\ell = v/R$, reducing the dynamics to passive (or usual) Brownian motion in the limit $\ell\downarrow 0$.  Secondly, when $D\chi= R$ we get $T=T_\text{eff}$ and detailed balance gets restored, with equilibrium density (for $U$ confining)
\[
\rho(x,y;\theta) \simeq e^{-\beta\, E(x,y;\theta)},\qquad \beta^{-1} = k_BT = k_BT_\text{eff}
\]
We can proceed from this case, by varying $\chi$, and basically repeating the
same analysis as for run-and-tumble particles.  The first dynamical phase is when $\chi \leq 0$, where the system is unconfined when $U=0$  see \fig \ref{2dn}, and the second phase happens for $\chi> 0$ where the dynamics is confined, even when $U=0$, see \fig \ref{2dp}.  Furthermore, the notion of heat and the characterization of the close-to-equilibrium regime is clear, as in the previous sections for the RTPs, since local detailed balance is verified for all $\chi$.

\section{Conclusions}
The embedding of active particle models in two-temperature processes allows the use of statistical thermodynamics applied to small systems.  Local detailed balance gets restored, and a clearer prescription of heat and thermal properties becomes possible.\\
We have illustrated the general procedure using simple models of bacterial locomotion (run-and-tumble and active Brownian particles), and by providing examples of hot spots (molecular switches and agitated vibrational modes).  The two-temperature embedding of the run-and-tumble process has in addition shown the robustness of the shape transition between a more confined regime and the occurrence of edge states, as monitored by the tumbling rate. \\
\vspace{1cm}

\noindent{\bf Acknowledgment} we thank Hidde Van Wiechen (TU Delft) for discussions leading to the solution presented in Fig. \ref{hidden}.
\newpage

\bibliographystyle{unsrt}  %ieeetr
\bibliography{chr}
\onecolumngrid
\end{document}